\documentclass[prd,aps,showpacs,epsf,floats,amsfonts,amssymb,amsmath,nofootinbib]{revtex4}

\newcommand{\be}{\begin{equation}}\newcommand{\ee}{\end{equation}}
\newcommand{\bea}{\begin{eqnarray} }\newcommand{\eea}{\end{eqnarray}}
\newcommand{\beaa}{\begin{eqnarray}}\newcommand{\eeaa}{\end{eqnarray}}
\newcommand{\ba}{\begin{array}}\newcommand{\ea}{\end{array}}
\newcommand{\bit}{\begin{itemize}}\newcommand{\eit}{\end{itemize}}
\newcommand{\ben}{\begin{enumerate}}\newcommand{\een}{\end{enumerate}}

\def\lan{\langle}
\def\lf{\left}
\def\non{\nonumber}
\def\ran{\rangle}
\def\ri{\right}

\def\al{\alpha}\def\bt{\beta}

\def\te{\theta}

\def\lan{\langle}
\def\lf{\left}

\def\non{\nonumber}
\def\ran{\rangle}

\def\ri{\right}

\def\al{\alpha}\def\bt{\beta}

\def\te{\theta}

\def\1{{_{1}}}\def\2{{_{2}}}

\begin{document}

\title{Dark energy induced by neutrino mixing}

\author{A.Capolupo${}^{\flat}$, S.Capozziello${}^{\sharp}$, G.Vitiello${}^{\flat}$}

\vspace{2mm}

\address{${}^{\flat}$Dipartimento di  Fisica "E.R. Caianiello"
and INFN, Universit\`a di Salerno, I-84100 Salerno, Italy
\\ ${}^{\sharp}$ Dipartimento di Scienze Fisiche, Universit\`a di Napoli "Federico II" and INFN Sez. di Napoli,
Compl. Univ. Monte S. Angelo, Ed.N, Via Cinthia, I-80126 Napoli,
Italy}

\begin{abstract}

The energy content of the vacuum condensate induced by the
neutrino mixing is interpreted as dynamically evolving dark
energy.

\end{abstract}

\maketitle

\section{Introduction}

The neutrino-antineutrino pair condensate in the vacuum can
provide a  contribution to the cosmological dark energy without
invoking {\it ad hoc} scalar field or further exotic mechanisms
\cite{Capolupo:2006et}. This result is obtained by analyzing the
neutrino mixing \cite{Pontecorvo:1957cp}-- \cite{Bilenky}
 in the context of Quantum Field Theory (QFT)
 \cite{BV95}--\cite{Blasone:2006jx}. In fact,
the unitary inequivalence has been recognized between the flavored
vacuum and the massive neutrino vacuum \cite{BV95,BHV98}. The
non--perturbative vacuum structure associated with the field
mixing \cite{BV95}--\cite{Blasone:2006jx} leads to new oscillation
formulas \cite{BHV98}-- \cite{yBCV02} (similar results hold in the
boson mixing case \cite{BCRV01,Capolupo:2004pt}) and to a
contribution to the vacuum energy
\cite{Capolupo:2006et,Blasone:2004yh}.

The vacuum condensate  can give rise to a dynamical (unclustered)
source of energy contributing to the today observed acceleration
of Hubble flow. Indeed, it provides a contribution
$\rho_{vac}^{mix}$, which in the early universe satisfies the
strong energy condition (SEC) $\rho_{vac}^{mix} + 3 p_{vac}^{mix}
\geq 0$, and behaves approximatively as a cosmological constant at
present epoch. Here $p_{vac}^{mix}$ is the vacuum pressure induced
by the neutrino mixing.
 A value of $\rho^{mix}_{vac}$, which is in agreement with
the estimated value of the dark energy, may be obtained by using a
cut-off on the momenta phenomenologically relevant for the
neutrino mixing phenomenon.

In Sect.II, we outline the neutrino mixing formalism in Quantum
Field Theory. In Sect.III we compute the mixing contribution to
the dark energy and the conclusions are drawn in Sect.IV.

\section{Neutrino mixing in Quantum Field Theory}

The main features of the QFT formalism for the neutrino mixing are
summarized as follows. For simplicity we restrict ourselves to the
two flavor case \cite{Blasone:2005ae}. Extension to three flavors
can be found in ref. \cite{yBCV02}, and similar results hold in
the boson mixing case \cite{BCRV01,Capolupo:2004pt,JM01,JM011},
(for a detailed review see \cite{Capolupo:2004av}).

We consider two Dirac neutrino fields. The Pontecorvo mixing
transformations are \cite{Pontecorvo:1957cp}
\begin{eqnarray} \nonumber\label{mix}
\nu_{e}(x) &=&\nu_{1}(x)\,\cos\theta + \nu_{2}(x)\,\sin\theta
\\
\nu_{\mu}(x) &=&-\nu_{1}(x)\,\sin\theta + \nu_{2}(x)\,\cos\theta
\;,\end{eqnarray}
where $\nu_{e}(x)$ and $\nu_{\mu}(x)$ are the fields with definite
flavors, $\theta$ is the mixing angle and $\nu_1$ and $\nu_2$ are
the fields with definite masses $m_{1} \neq m_{2}$. $\nu_{1}(x)$
and $\nu_{2}(x)$ are
 \bea\label{freefi}
 \nu _{i}(x)=\frac{1}{\sqrt{V}}{\sum_{{\bf k} ,
r}} \left[ u^{r}_{{\bf k},i}\, \al^{r}_{{\bf k},i}(t) +
v^{r}_{-{\bf k},i}\, \bt^{r\dag}_{-{\bf k},i}(t) \ri] e^{i {\bf
k}\cdot{\bf x}}, \eea with $i=1,2$, $ \al_{{\bf
k},i}^{r}(t)=\al_{{\bf k},i}^{r}\, e^{-i\omega _{k,i}t}$, $
\bt_{{\bf k},i}^{r\dag}(t) = \bt_{{\bf k},i}^{r\dag}\,
e^{i\omega_{k,i}t},$ and $ \omega _{k,i}=\sqrt{{\bf k}^{2} +
m_{i}^{2}}.$

The operators $\alpha ^{r}_{{\bf k},i}$ and $ \beta ^{r }_{{\bf
k},i}$, $ i=1,2 \;, \;r=1,2$ are the annihilators for the vacuum
state $|0\rangle_{1,2}\equiv|0\rangle_{1}\otimes |0\rangle_{2}$:
$\alpha ^{r}_{{\bf k},i}|0\rangle_{12}= \beta ^{r }_{{\bf
k},i}|0\rangle_{12}=0$.

The mixing transformation Eqs.(\ref{mix}) can be written as
\cite{BV95}:
\bea \label{mixG} \nu_{e}^{\alpha}(x) = G^{-1}_{\bf \te}(t)\;
\nu_{1}^{\alpha}(x)\; G_{\bf \te}(t) \\ \non \nu_{\mu}^{\alpha}(x)
= G^{-1}_{\bf \te}(t)\; \nu_{2}^{\alpha}(x)\; G_{\bf \te}(t) \eea
where $G_{\bf \te}(t)$ is the mixing generator.

 At finite volume $G_{\bf \te}(t)$ is an unitary
operator: $G^{-1}_{\bf \te}(t)=G_{\bf -\te}(t)=G^{\dag}_{\bf
\te}(t)$, preserving the canonical anticommutation relations,
moreover $G^{-1}_{\bf \te}(t)$ maps the Hilbert spaces for free
fields ${\cal H}_{1,2}$ to the Hilbert spaces for interacting
fields ${\cal H}_{e,\mu}$: $ G^{-1}_{\bf \te}(t): {\cal H}_{1,2}
\mapsto {\cal H}_{e,\mu}.$ In particular for the vacuum $|0
\rangle_{1,2}$ we have, at finite volume: $ |0(t) \rangle_{e,\mu}
= G^{-1}_{\bf \te}(t)\; |0 \rangle_{1,2}\;. $ $|0 \rangle_{e,\mu}$
is the vacuum for ${\cal H}_{e,\mu}$, which we will refer to as
the flavor vacuum.

The flavor annihilators, relative to the fields $\nu_{e}(x)$ and
$\nu_{\mu}(x)$ are defined as (we use $(\sigma,i)=(e,1) ,
(\mu,2)$):
 \begin{eqnarray}\label{flavannich}
\alpha _{{\bf k},\sigma}^{r}(t) &\equiv &G^{-1}_{\bf
\te}(t)\;\alpha _{{\bf k},i}^{r}(t)\;G_{\bf \te}(t),  \nonumber
\\
\beta _{{\bf k},\sigma}^{r}(t) &\equiv &G^{-1}_{\bf \te}(t)\;\beta
_{{\bf k},i}^{r}(t)\;G_{\bf \te}(t).
\end{eqnarray}

The flavor fields can be then expanded in the same bases as
$\nu_{i}$:
\begin{eqnarray}
\nu _{\sigma}({\bf x},t) &=&\frac{1}{\sqrt{V}}{\sum_{{\bf k},r} }
e^{i{\bf k.x}}\left[ u_{{\bf k},i}^{r} \alpha _{{\bf
k},\sigma}^{r}(t) + v_{-{\bf k},i}^{r} \beta _{-{\bf
k},\sigma}^{r\dagger }(t)\right] .
\end{eqnarray}

The flavor annihilation operators in the reference frame such that
${\bf k}=(0,0,|{\bf k}|)$ are explicitely \bea\label{annihilator}
\alpha^{r}_{{\bf k},e}(t)&=&\cos\theta\;\alpha^{r}_{{\bf
k},1}(t)\;+\;\sin\theta\;\left( |U_{{\bf k}}|\; \alpha^{r}_{{\bf
k},2}(t)\;+\;\epsilon^{r}\; |V_{{\bf k}}|\; \beta^{r\dag}_{-{\bf
k},2}(t)\right) \eea (and similar for $\alpha^{r}_{{\bf k},\mu}$,
$\beta^{r}_{-{\bf k},e}$,
 $\beta^{r}_{-{\bf k},\mu}$), with $\epsilon^{r}=(-1)^{r}$ and
\bea\label{Vk2}
 |U_{{\bf k}}| \equiv u^{r\dag}_{{\bf k},i}
u^{r}_{{\bf k},j} = v^{r\dag}_{-{\bf k},i} v^{r}_{-{\bf k},j}\,,
\qquad |V_{{\bf k}}| \equiv  \epsilon^{r}\; u^{r\dag}_{{\bf k},1}
v^{r}_{-{\bf k},2} = -\epsilon^{r}\; u^{r\dag}_{{\bf k},2}
v^{r}_{-{\bf k},1} \eea with $i,j = 1,2$ and $ i \neq j$. We have:
  \bea
\non |U_{{\bf
k}}|=\left(\frac{\omega_{k,1}+m_{1}}{2\omega_{k,1}}\right)^{\frac{1}{2}}
\left(\frac{\omega_{k,2}+m_{2}}{2\omega_{k,2}}\right)^{\frac{1}{2}}
\left(1+\frac{{\bf
k}^{2}}{(\omega_{k,1}+m_{1})(\omega_{k,2}+m_{2})}\right)
\\
\label{Vk}|V_{{\bf
k}}|=\left(\frac{\omega_{k,1}+m_{1}}{2\omega_{k,1}}\right)^{\frac{1}{2}}
\left(\frac{\omega_{k,2}+m_{2}}{2\omega_{k,2}}\right)^{\frac{1}{2}}
\left(\frac{k}{(\omega_{k,2}+m_{2})}-\frac{k}{(\omega_{k,1}+m_{1})}\right)
\eea \bea |U_{{\bf k}}|^{2}+|V_{{\bf k}}|^{2}=1. \eea

The condensation density is given by
\bea\label{condensation}
 _{e,\mu}\langle 0| \al_{{\bf k},i}^{r \dag} \al^r_{{\bf k},i}
|0\rangle_{e,\mu}\,= \;_{e,\mu}\langle 0| \bt_{{\bf k},i}^{r \dag}
\bt^r_{{\bf k},i} |0\rangle_{e,\mu}\,=\, \sin^{2}\te\; |V_{{\bf
k}}|^{2} \;, \qquad i=1,2\,. \eea

\section{Neutrino mixing and dark energy}

Experimental data \cite{CMBR}--\cite{SNeIa} support the picture
that some form of {\it dark energy}, evolving from early epochs,
induces the today observed acceleration of the universe.

There are many proposals to achieve cosmological models justifying
such a dark component
\cite{Capozziello:2005ku}--\cite{Deffayet:2001pu}. In this
Section, we present the contribution $\rho_{vac}^{mix}$ of the
neutrino mixing to the vacuum energy density. We consider the
Minkowski metric.

The Lorentz invariance of the vacuum implies that the vacuum state
is the zero eigenvalue eigenstate of the normal ordered energy,
momentum and angular momentum operators \cite{Itz}. Therefore
${\cal T}_{\mu\nu}^{vac} = \lan 0 |:{\cal T}_{\mu\nu}:| 0\ran= 0$.
The (0,0) component of the energy-momentum tensor density
 ${\cal T}_{00}(x) $ for the fields $\nu_1$ and $\nu_2$ is then
\bea\
 :{\cal T}_{00}(x): = \frac{i}{2}:\left({\bar \Psi}_{m}(x)\gamma_{0}
\stackrel{\leftrightarrow}{\partial}_{0} \Psi_{m}(x)\right): \eea
where $:...:$ denotes  the customary normal ordering with respect
to the mass vacuum in the flat space-time and $\Psi_{m} = (\nu_1,
\nu_2)$.

In terms of the annihilation and creation operators of fields
$\nu_{1}$ and $\nu_{2}$, the (0,0) component of the
energy-momentum tensor $ T_{00}=\int d^{3}x {\cal T}_{00}(x)$ is
given by
\bea\label{T00}
 :T^{00}_{(i)}:= \sum_{r}\int d^{3}{\bf k}\,
\omega_{k,i}\lf(\al_{{\bf k},i}^{r\dag} \al_{{\bf k},i}^{r}+
\beta_{{\bf -k},i}^{r\dag}\beta_{{\bf -k},i}^{r}\ri), \eea with
$i=1,2$. Note that $T^{00}_{(i)}$ is time independent.

The contribution $\rho_{vac}^{mix}$ of the neutrino mixing to the
vacuum energy density is given by computing the expectation value
of $T^{00}_{(i)}$ in the flavor vacuum $|0 {\rangle}_{e,\mu}$:
 \bea\
\rho_{vac}^{mix} = \frac{1}{V}\; \eta_{00}\; {}_{e,\mu}\lan 0(t)
|\sum_{i} :T^{00}_{(i)}(0):| 0(t)\ran_{e,\mu}   ~.
 \eea

Within the QFT formalism for neutrino mixing, we have
 \bea
 {}_{e,\mu}\lan 0 |:T^{00}_{(i)}:| 0\ran_{e,\mu}={}_{e,\mu}\lan
0(t) |:T^{00}_{(i)}:| 0(t)\ran_{e,\mu}
 \eea
  for any t.
 We then obtain
\bea\non \rho_{vac}^{mix} &= &\sum_{i,r}\int \frac{d^{3}{\bf
k}}{(2 \pi)^{3}} \, \omega_{k,i}\Big({}_{e,\mu}\lan 0 |\al_{{\bf
k},i}^{r\dag} \al_{{\bf k},i}^{r}| 0\ran_{e,\mu} + {}_{e,\mu}\lan
0 |\beta_{{\bf k},i}^{r\dag} \beta_{{\bf k},i}^{r}| 0\ran_{e,\mu}
\Big) . \eea
By using Eq.(\ref{condensation}), we get \bea\label{cc}
 \rho_{vac}^{mix} = \frac{ 2}{\pi} \sin^{2}\theta
\int_{0}^{K} dk \, k^{2}(\omega_{k,1}+\omega_{k,2}) |V_{\bf
k}|^{2} , \eea
where the cut-off $K$ has been introduced.

In a similar way, the contribution
 $ p_{vac}^{mix}$ of the neutrino mixing
to the vacuum pressure is given by the expectation value of
$T^{jj}_{(i)}$ (where no summation on the index $j$ is intended)
on the flavor vacuum $| 0\ran_{e,\mu}$:
 \bea\
p_{vac}^{mix}= -\frac{1}{V}\; \eta_{jj} \; {}_{e,\mu}\lan 0
|\sum_{i} :T^{jj}_{(i)}:| 0\ran_{e,\mu} ~.
 \eea
where no summation on the index $j$ is intended. Being
 \bea\label{Tjj}
 :T^{jj}_{(i)}:= \sum_{r}\int d^{3}{\bf k}\, \frac{k^j
k^j}{\;\omega_{k,i}}\lf(\al_{{\bf k},i}^{r\dag} \al_{{\bf
k},i}^{r}+ \beta_{{\bf -k},i}^{r\dag}\beta_{{\bf -k},i}^{r}\ri),
\eea
 in the case of the
isotropy of the momenta we have $T^{11} = T^{22} = T^{33}$, then
 \bea\label{cc2}
  p_{vac}^{mix} = \frac{2}{3\;\pi}
\sin^{2}\theta \int_{0}^{K} dk \,
k^{4}\lf[\frac{1}{\omega_{k,1}}+\frac{1}{\omega_{k,2}}\ri] |V_{\bf
k}|^{2}\,.
 \eea
By Eqs.(\ref{cc}) and (\ref{cc2}) we have that Lorentz invariance
of vacuum neutrino-antineutrino condensate is broken, indeed
 $\rho_{vac}^{mix} \neq - p_{vac}^{mix}$ for any
value of $m_1$ and $m_2$ and  {\it independently of the choice} of
the cut-off. We note that the adiabatic index $w = p_{vac}^{mix}/
\rho_{vac}^{mix} \simeq 1/3$ when the cut-off is chosen to be $K
\gg m_{1}, m_{2}$.

The values of $\rho_{vac}^{mix}$ and  $ p_{vac}^{mix}$ we obtain
are time-independent since, for the sake of simplicity, we are
taking into account the Minkowski metric. When the curved
background metric is considered, $|V_{\bf k}|^{2}$ is
time-dependent. Such a result holds in the early universe epochs,
when the  curvature radius is comparable with the oscillation
length.

At the present epoch, the breaking of the Lorentz invariance is
negligible and then
 $\rho_{vac}^{mix}$ comes from space-time independent condensate contributions
 (i.e. the contributions carrying a non-vanishing $\partial_{\mu} \sim k_{\mu}=(\omega_{k},k_{j}) $
are missing).
 This  means that the stress energy tensor of the vacuum condensate
is  approximatively equal to \bea {}_{e,\mu}\lan 0 |:T_{\mu\nu}:|
0\ran_{e,\mu} = \eta_{\mu\nu}\;\sum_{i}m_{i}\int
\frac{d^{3}x}{(2\pi)^3}\;{}_{e,\mu}\lan 0 |:\bar{\nu
}_{i}(x)\nu_{i}(x):| 0\ran_{e,\mu}\, =
\eta_{\mu\nu}\;\rho_{\Lambda}^{mix}.
 \eea

Since $\eta_{\mu\nu} = diag (1,-1,-1,-1)$ and, in a homogeneous
and isotropic universe, the energy momentum tensor is $T_{\mu\nu}
= diag (\rho\,,p\,,p\,,p\,)$, then, consistently with Lorentz
invariance, the state equation is $\rho_{\Lambda}^{mix} =
-p_{\Lambda}^{mix}$.

That is, today the vacuum condensate coming from the neutrino
mixing, contributes to the dynamics of the universe, with a
behavior similar to that of the cosmological constant
{\cite{Blasone:2004yh}. Explicitly, we have \bea \label{cost}
\rho_{\Lambda}^{mix}=  \frac{2}{\pi} \sin^{2}\theta \int_{0}^{K}
dk \, k^{2}\lf[\frac{m_{1}^{2}}{\omega_{k,1}}+\frac{m_{2}^{2}}
{\omega_{k,2}}\ri] |V_{\bf k}|^{2}.
 \eea

It is shown that during the Big Bang Nucleosyntesis, flavor
particle-antiparticle pairs are produced by mixing and
oscillations with typical momentum $k \sim \frac{m_{1} +
m_{2}}{2}$ \cite{Boyanovsky:2004xz}. This introduces us to comment
on the problem of the choice of the cut-off $K$ in the
integrations in the equations above.

We do not have the solution for such a problem. However, in our
approach there is an indication of possible choices suggested by
the natural energy scale of the neutrino mixing in the QFT
formalism. The ultraviolet cut-off at Planck scale, as well as the
QCD one, give huge unacceptable values for the today observed
vacuum energy density. It is therefore imperative to explore
alternative routes. One of the merits of the present approach is
indeed to point out that, although the arbitrariness problem is
not solved, other possible choices exist which not only are
consistent with the intrinsic energy scale of the mixing
phenomenon, but also lead to quite acceptable values for the
vacuum energy density. We thus arrive at the cut-off choice which
is suggested by the natural scale appearing in the QFT formalism
of the mixing phenomenon, i.e. we may set $K\simeq \sqrt{m_{1}
m_{2}}$ \cite{Blasone:2004yh}. Another possibility, as suggested
in Ref. \cite{Barenboim:2004ev} on similar grounds, is the sum of
the two neutrino masses, $K = m_{1} + m_{2}$. Both choices lead to
values of $\rho_{\Lambda}^{mix}$ compatible with the observed
value of $\rho_{\Lambda}$. The latter choice is also quite near to
another possibility, $K \sim \frac{m_{1} + m_{2}}{2}$, which could
be related to the discussion of Ref. \cite{Boyanovsky:2004xz},
although this is referred to background neutrinos. It is indeed an
interesting open question  the relation between the (hot) dark
matter and the dark energy, namely, from the perspective of the
present paper, of the relation between dark matter and the vacuum
structure. Such a question will be object of our future study.

By using $\sin^{2}\theta\simeq 0.3$,  $m_{i}$ of the order of $
10^{-3}eV$, so that $\delta m^{2}=m_{2}^{2}-m_{1}^{2}\simeq 8
\times 10^{-5}eV^{2}$, and one of the above choices for $K$, for
example $K \sim \frac{m_{1} + m_{2}}{2}$, we obtain
$\rho_{\Lambda}^{mix} \sim 1.1 \times 10^{-47}GeV^{4}$,
 which is in agreement with the estimated value of
the dark energy. The other two choices lead to values of
$\rho_{\Lambda}^{mix}$ also compatible with the estimated value of
$\rho_{\Lambda}$, i.e. $\rho_{\Lambda}^{mix} \sim 0.7 \times
10^{-47}GeV^{4}$ and $\rho_{\Lambda}^{mix} \sim 5.5 \times
10^{-47}GeV^{4}$, respectively.

We remark that, unless one works in the present approach, such
incredibly small values for the cut-off would be ruled out, since
one would think that regularization of quantum effects from the
physics beyond the standard model should come at a very high
scale, e. g. the Planck scale. However, in the present case such a
belief is actually unfounded: it is indeed in conflict with simple
facts such as the disagreement of 123 orders of magnitude with the
observed dark energy value. On the contrary, being bounded to a
flat computational basis, as observed in Refs.
\cite{Blasone:2004yh,Barenboim:2004ev}, the presence of $|V_{\bf
k}|^{2}$ (with its behavior as a function of the momentum) in the
integrations naturally leads to one of the above small cut-off
choices. The non-perturbative physics of the neutrino mixing thus
points to the relevance of soft momentum (long-wave-length) modes.

In this connection, we also remark that Eqs.(\ref{cc}) and
(\ref{cost}) show that the contribution to the dark energy induced
from the neutrino mixing depends on the specific QFT nature of the
mixing: indeed, it is absent in the quantum mechanical treatment
of the mixing.

\section{Conclusions}

We have reported on recent results showing that the vacuum
condensate due to neutrino mixing contributes to the dark energy
of the universe. The expectation value of the energy-momentum
tensor has been computed in the vacuum state where neutrino
oscillations are observed and the energy content of the vacuum
condensate induced by the neutrino mixing is interpreted as
dynamically evolving dark energy. By careful choice of the
momentum cut-off we have obtained acceptable values for vacuum
energy density.


\medskip

\section*{References}

\bibliography{apssamp}

\begin{thebibliography}{99}


\bibitem{Capolupo:2006et}
  A.~Capolupo, S.~Capozziello and G.~Vitiello,
  Phys. Lett. A in print, arXiv:astro-ph/0602467 (2006).

\bibitem{Pontecorvo:1957cp}
  B.~Pontecorvo,
  Sov.\ Phys.\ JETP {\bf 6}, 429 (1957)
  [Zh.\ Eksp.\ Teor.\ Fiz.\  {\bf 33}, 549 (1957)].

\bibitem{Bilenky:1978nj}
  S.~M.~Bilenky and B.~Pontecorvo,
  Phys.\ Rept.\  {\bf 41}, 225 (1978).

\bibitem{Bilenky} S.M. Bilenky and
S.T. Petcov,
Rev.\ Mod.\ Phys.\ {\bf 59}, 671 (1987).
\\
T.~Cheng and L.~Li, {\it Gauge Theory of Elementary Particle
Physics}, Clarendon Press, Oxford, 1989.

\bibitem{BV95}
M. Blasone and G. Vitiello,
Annals Phys.\ {\bf 244}, 283 (1995).

\bibitem{BHV98}
M. Blasone, P.A. Henning and G. Vitiello, Phys.\ Lett.\ B {\bf
451},140 (1999).

\bibitem{Fujii:1999xa}
K.~Fujii, C.~Habe and T.~Yabuki, Phys.\ Rev.\ D {\bf 59}, 113003
(1999);
 Phys.\ Rev.\ D {\bf 64}, 013011 (2001).

\bibitem{JM01}
  C.~R.~Ji and Y.~Mishchenko,
  Phys.\ Rev.\ D {\bf 64}, 076004 (2001).

\bibitem{JM011}
  C.~R.~Ji and Y.~Mishchenko,
  Phys.\ Rev.\ D {\bf 65}, 096015 (2002).

\bibitem{yBCV02}
M.~Blasone, A.~Capolupo and G.~Vitiello,
Phys.\ Rev.\ D {\bf 66}, 025033 (2002).

\bibitem{Capolupo:2004av}
A.~Capolupo, Ph.D. Thesis
[hep-th/0408228].

\bibitem{Blasone:2005ae}
M.~Blasone, A.~Capolupo, F.~Terranova and G.~Vitiello,
Phys.\ Rev.\ D {\bf 72}013003 (2005).

\bibitem{Blasone:2005tk}
  M.~Blasone, A.~Capolupo and G.~Vitiello,
  Acta Phys.\ Polon.\ B {\bf 36} 3245 (2005).

\bibitem{Blasone:2006jx}
  M.~Blasone, A.~Capolupo, C.~R.~Ji and G.~Vitiello,
  hep-ph/0611106 (2006).


\bibitem{BCRV01}
  M.~Blasone, A.~Capolupo, O.~Romei and G.~Vitiello,
  Phys.\ Rev.\ D {\bf 63}, 125015 (2001).


\bibitem{Capolupo:2004pt}
A.~Capolupo, C.~R.~Ji, Y.~Mishchenko and G.~Vitiello,
Phys.\ Lett.\ B {\bf 594}, 135 (2004).


\bibitem{Blasone:2004yh}
  M.~Blasone, A.~Capolupo, S.~Capozziello, S.~Carloni and G.~Vitiello,
  Phys.\ Lett.\ A {\bf 323}  182 (2004).

\bibitem{CMBR}
P. de Bernardis et al., Nature, 404, 955, 2000;
\\ D.N. Spergel et
al. ApJS, 148, 175 (2003).

\bibitem{LSS}
S. Dodelson et al., ApJ, 572, 140 (2002).
\\ W.J. Percival et al., MNRAS, 337, 1068 (2002).
\\  A.S. Szalay et al., ApJ, 591, 1, (2003).
\\  E. Hawkins et al., MNRAS, 346, 78 (2003).
\\  A.C. Pope et al., ApJ, 607,
655 (2004).

\bibitem{SNeIa}
 A.G. Riess et al., ApJ, 607, 665 (2004).

\bibitem{Capozziello:2005ku}
R. Kerner, Gen. Rel. Grav. {\bf 14}, 453 (1982).
\\
  S.~Capozziello, V.~F.~Cardone and A.~Troisi,
  Phys.\ Rev.\ D {\bf 71}, 043503 (2005).
  \\
  T.P. Sotiriou, Phys.\ Rev.\ D {\bf 73}, 063515 (2006).


\bibitem{Sahni:1999gb}
  V.~Sahni and A.~A.~Starobinsky,
  Int.\ J.\ Mod.\ Phys.\ D {\bf 9}, 373 (2000).

\bibitem{Deffayet:2001pu}
  C.~Deffayet, G.~R.~Dvali and G.~Gabadadze,
  Phys.\ Rev.\ D {\bf 65}, 044023 (2002).



\bibitem{Itz}
C.~Itzykson and J.~B.~Zuber, {\it Quantum Field Theory},
(McGraw-Hill, New York, 1980).
\\
S. Schweber, {\it An itroduction Relativistic Quantum Field
Theory}, (Harper and Row, 1961).

\bibitem{Boyanovsky:2004xz}
  D.~Boyanovsky and C.~M.~Ho,
  Phys.\ Rev.\ D {\bf 69}, 125012 (2004).

\bibitem{Barenboim:2004ev}
  G.~Barenboim and N.~E.~Mavromatos,
  Phys.\ Rev.\ D {\bf 70}, 093015 (2004).


\end{thebibliography}

\end{document}